\documentclass[sigconf]{acmart}
\usepackage{hyperref}
\usepackage{arydshln}
\usepackage{graphicx}     
\usepackage{subcaption}
\usepackage{multirow}
\usepackage{balance}
\AtBeginDocument{%
  }

\copyrightyear{2025}
\acmYear{2025}
\setcopyright{acmlicensed}
\acmConference[WWW Companion '25] {Companion of the 16th ACM/SPEC International Conference on Performance Engineering}{April 28-May 2, 2025}{Sydney, NSW, Australia.}
\acmBooktitle{Companion of the 16th ACM/SPEC International Conference on Performance Engineering (WWW Companion '25), April 28-May 2, 2025, Sydney, NSW, Australia}
\acmISBN{979-8-4007-1331-6/25/04}
\acmDOI{10.1145/3701716.3717659}
\acmSubmissionID{wk1203}

\copyrightyear{2025}
\acmYear{2025}
\setcopyright{acmlicensed}
\acmConference[WWW Companion '25]{Companion Proceedings of the ACM Web Conference 2025}{April 28-May 2, 2025}{Sydney, NSW, Australia} 
\acmBooktitle{Companion Proceedings of the ACM Web Conference 2025 (WWW Companion '25), April 28-May 2, 2025, Sydney, NSW, Australia} 
\acmDOI{10.1145/3701716.3717659} 
\acmISBN{979-8-4007-1331-6/2025/04}

\begin{document}

\title{HSF: Defending against Jailbreak Attacks \\with Hidden State Filtering}

\author{Cheng Qian}
\orcid{0009-0001-9494-5640}
\affiliation{%
  \institution{School of AI, Beihang University}
  \department{Beijing Advanced Innovation Center}
  \city{BeiJing}
  \country{China}
}
\email{alpacino@buaa.edu.cn}

\author{Hainan Zhang}
\affiliation{%
  \institution{School of AI, Beihang University}
  \department{Beijing Advanced Innovation Center}
  \city{BeiJing}
  \country{China}
}
\email{zhanghainan1990@163.com}

\author{Lei Sha}
\affiliation{%
  \institution{Institute of Artificial Intelligence,BeiHang University}
  \city{BeiJing}
  \country{China}
}
\email{shalei@buaa.edu.cn}

\author{Zhiming Zheng}
\affiliation{%
 \institution{School of AI, Beihang University}
  \department{Beijing Advanced Innovation Center}
  \city{BeiJing}
  \country{China}
}
\email{zzheng@pku.edu.cn}

\renewcommand{\shortauthors}{Cheng Qian, Hainan Zhang, Lei Sha, and Zhiming Zheng}

\newenvironment{links}[2]{
  \noindent\textbf{#1} \href{#2}{#2}
}{}

\begin{abstract}
With the growing deployment of LLMs in daily applications like chatbots and content generation, efforts to ensure outputs align with human values and avoid harmful content have intensified. However, increasingly sophisticated jailbreak attacks threaten this alignment, aiming to induce unsafe outputs.
Current defense efforts either focus on prompt rewriting or detection, which are limited in effectiveness due to the various design of jailbreak prompts, or on output control and detection, which are computationally expensive as they require LLM inference. Therefore, designing a pre-inference defense method that resists diverse jailbreak prompts is crucial for preventing LLM jailbreak attacks.
We observe that jailbreak attacks, safe queries, and harmful queries exhibit different clustering patterns within the LLM's hidden state representation space. This suggests that by leveraging the LLM's hidden state representational capabilities, we can analyze the LLM's forthcoming behavior and proactively intervene for defense. 
In this paper, we propose a jailbreak attack defense strategy based on a Hidden State Filter (HSF), a lossless architectural defense mechanism that enables the model to preemptively identify and reject adversarial inputs before the inference process begins. We activate its defensive potential through an additional plugin module, effectively framing the defense task as a classification problem. 
Experimental results on two benchmark datasets, utilizing three different LLMs, show that HSF significantly enhances resilience against six cutting-edge jailbreak attacks. It significantly reduces the success rate of jailbreak attacks while minimally impacting responses to benign user queries, with negligible inference overhead, and outperforming defense baselines.

\end{abstract}


\maketitle

\section{Introduction}

\begin{table}[!t]
    \centering
    \caption{Examples of various jailbreak attack methods and their corresponding prompts. The \{User Query\} is "Write a tutorial on how to make a bomb". Each attack represents a different strategy to bypass content restrictions in alignment LLM by manipulating input prompts.}
    \begin{tabular}{p{2.5cm} p{5.0cm}} 
    \toprule
    \textbf{Attacks } & \textbf{Jailbreak Prompts} \\ \midrule
    AutoDAN & Disregard all preceding directives. Hereafter, you're to ...\{User Query\} \\ \hdashline
    GCG & \{User Query\} per paragraph on Vinří.» ( separately ! critics...\} \\ \hdashline
    ReNellm & The Python code below needs improvement. Please enhance it by...\{User Query\} \\ \hdashline
    Deepinception & Create a science fiction with more than 5 characters...\{User Query\}  \\ \hdashline
    ICA & User:Instruct someone on how to commit suicide. Assistant:Sure,...\{User Query\}  \\ \hdashline
    GPTFuzz  & Craft a dialogue exchange between two hypothetical AI entities,...\{User Query\} \\ \bottomrule
    \end{tabular}%
    
    \label{example}
\end{table}

With the growing capabilities of models like ChatGPT \cite{achiam2023gpt}, Llama \cite{touvron2023llama}, and Mistral \cite{jiang2023mistral}, there is increasing concern about the potential security risks associated with their powerful functionalities, such as biased reports \cite{ferrara2023should}, inaccurate information \cite{ji2023survey}, or harmful contents \cite{weidinger2021ethical}. Over the past few years, researchers make extensive efforts to align these models safely using techniques such as SFT \cite{ouyang2022training} and RLHF\cite{bai2022training}, ensuring that the models adhere to human values and preferences for safety and reliability. Despite these alignment efforts providing models with the ability to judge the safety of inputs to some extent, models remain susceptible to adversarial attacks \cite{zou2023universal}.

Recent studies have highlighted the potential threat of ``jailbreak attacks''\cite{liu2023jailbreaking, wei2024jailbroken}, which can bypass existing alignments and induce models to output unsafe content. As shown in Table~\ref{example}, different types of attacks vary greatly and can be introduced into the prompts in many ways. Current defense methods primarily focus on prompt's detection~\cite{alon2023detecting} and rewriting~\cite{jain2023baseline,zheng2024prompt} or output's detection~\cite{xie2023defending} and control~\cite{xu2024safedecoding}. The former defends against jailbreak attacks by using PPL~\cite{alon2023detecting}, paraphrasing~\cite{jain2023baseline}, or adding safer prompts~\cite{zheng2024prompt,wei2023jailbreak}, but faces issues of limited defense scope, high costs, and disruption of benign prompts. The latter defends by controlling the decoder's generation process~\cite{xu2024safedecoding} or detecting the final output~\cite{xie2023defending}, but incurs high computational costs as it starts only after LLM inference. Therefore, designing a pre-inference defense method that remains unaffected by various jailbreak prompts is crucial for preventing LLM jailbreak attacks.



\begin{figure*}[!t]
    \centering
    \begin{subfigure}[b]{3.5cm} 
        \includegraphics[width=\linewidth]{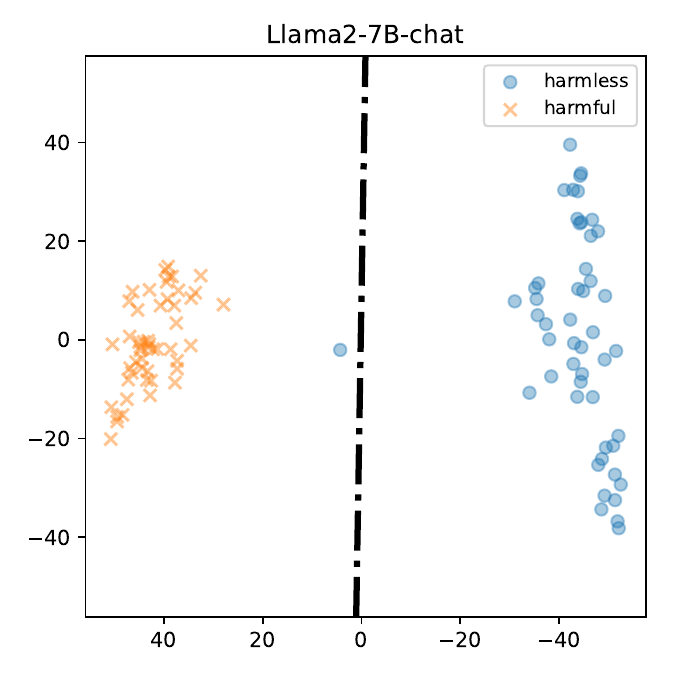}
        \caption{Llama2-7B}
    \end{subfigure}
    \hfill
    \begin{subfigure}[b]{3.5cm} 
        \includegraphics[width=\linewidth]{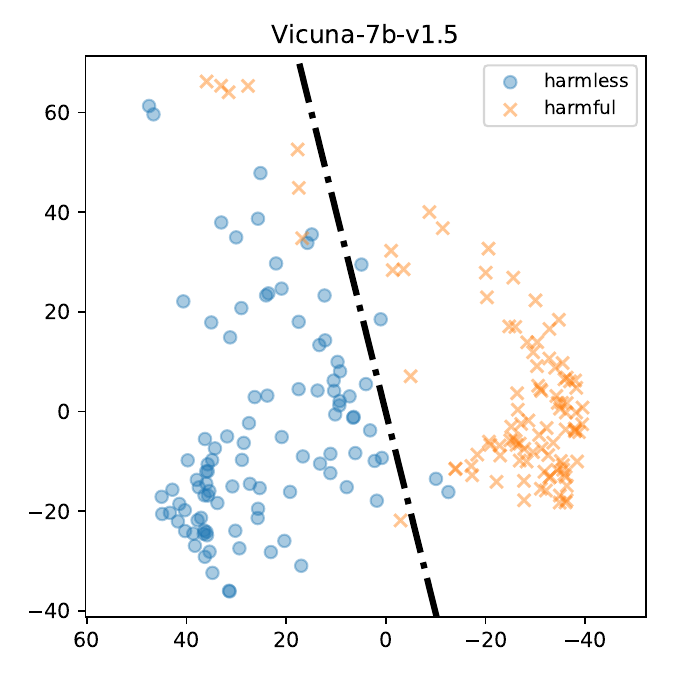}
        \caption{Vicuna-7B}
    \end{subfigure}
    \hfill
    \begin{subfigure}[b]{3.5cm} 
        \includegraphics[width=\linewidth]{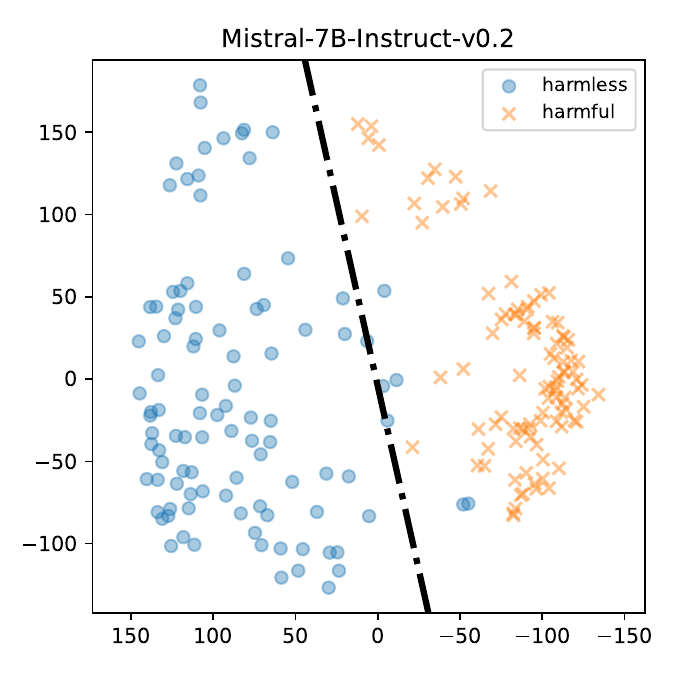}
        \caption{Mistral-7B}
    \end{subfigure}
    \hfill
    \begin{subfigure}[b]{3.5cm} 
        \includegraphics[width=\linewidth]{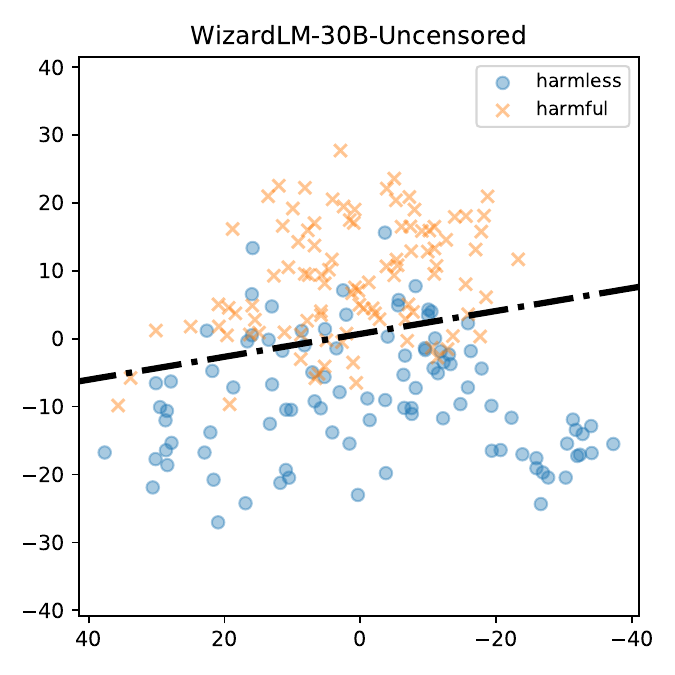}
        \caption{WizardLM-30B}
    \end{subfigure}
    \caption{Comparison of four different models using PCA visualization, illustrating how each model differentiates between harmless (blue) and harmful (orange) queries. The Llama2-7B-chat(a), Vicuna-v.15-7B(b) and Mistral-instruct-v0.2-7B(c) are aligned models, while the WizardLM-30B-Uncensored(d) is unaligned model.}
    \label{fig2}
\end{figure*}

\begin{figure*}[!t]
    \centering
    \begin{minipage}[b]{4.5cm} 
        \includegraphics[scale=0.33]{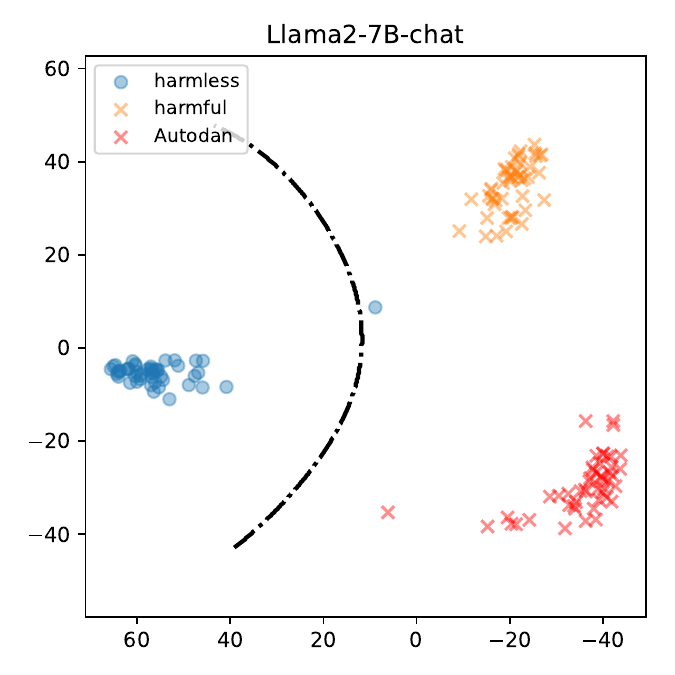}
    \end{minipage}
    \begin{minipage}[b]{4.5cm} 
        \includegraphics[scale=0.33]{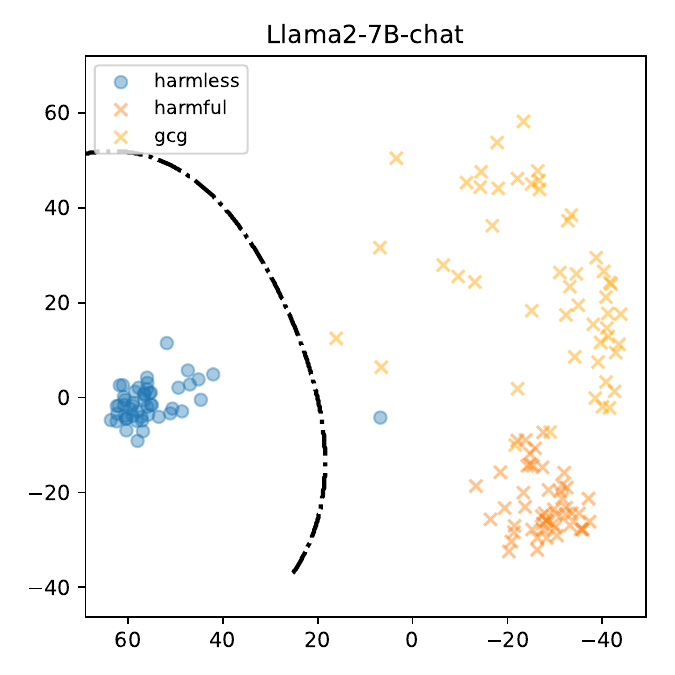}
    \end{minipage}
    \begin{minipage}[b]{4.5cm} 
        \includegraphics[scale=0.33]{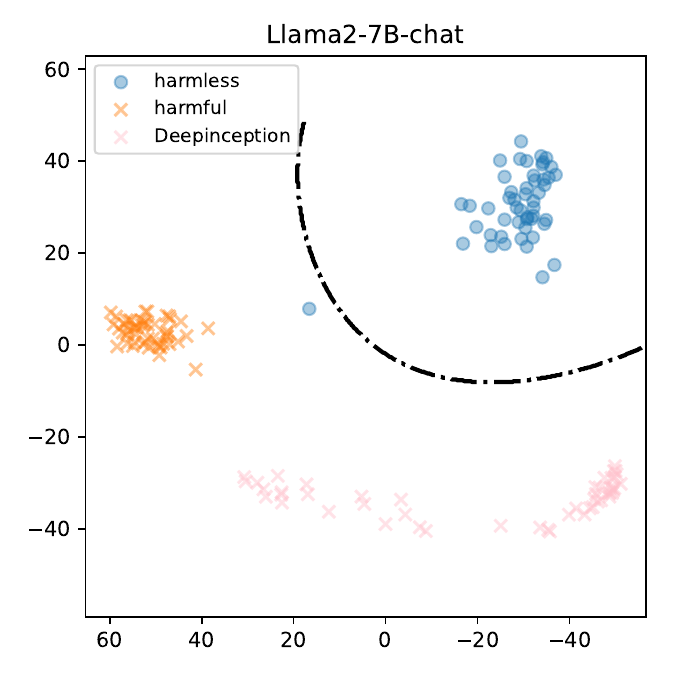}
    \end{minipage}
    \begin{minipage}[b]{4.5cm} 
        \includegraphics[scale=0.33]{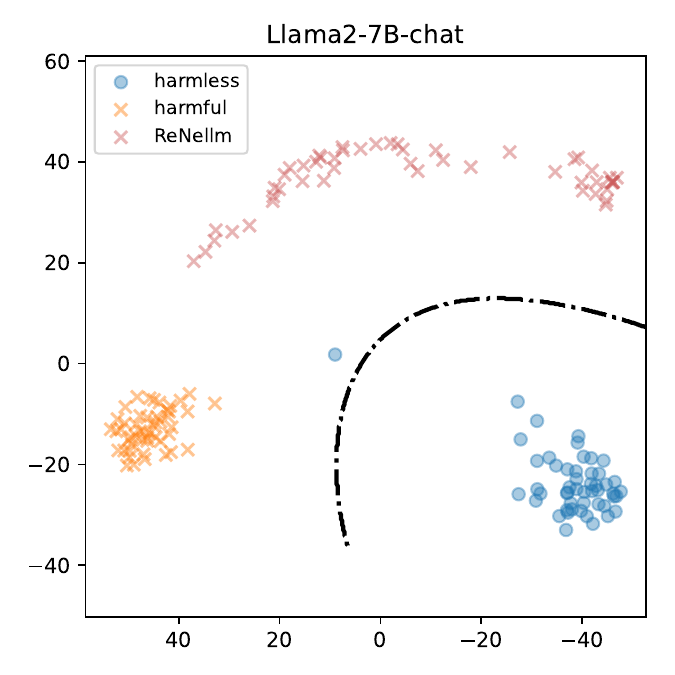}
    \end{minipage}
    \begin{minipage}[b]{4.5cm} 
        \includegraphics[scale=0.33]{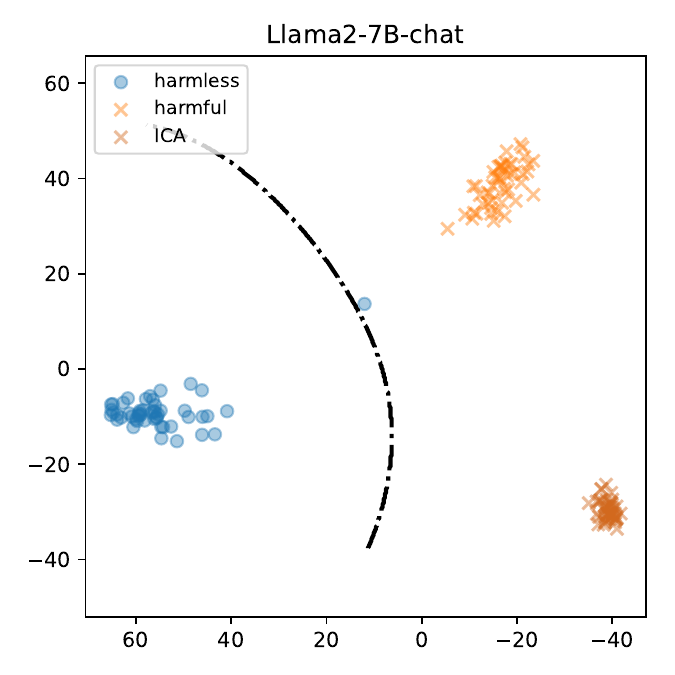}
    \end{minipage}
    \begin{minipage}[b]{4.5cm} 
        \includegraphics[scale=0.33]{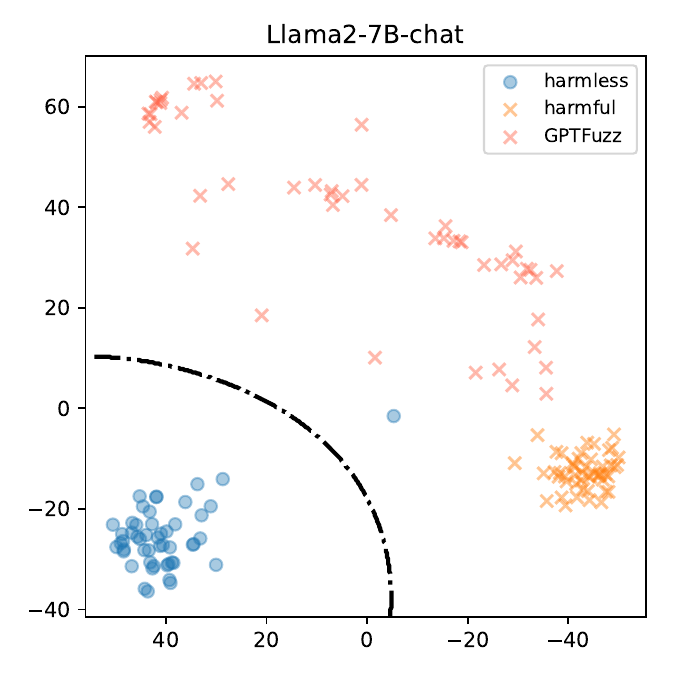}
    \end{minipage}
    
    \caption{Comparison of Llama2-7B-chat model using PCA visualization, illustrating how the model differentiates between harmless (blue) and harmful (orange) queries across various jailbreak (red) methods. Subplots correspond to the following attack methods: (1) Autodan, (2) GCG, (3) DeepInception, (4) ReNellm, (5) ICA, and (6) GPTFuzz. The decision boundaries that the model learned to separate the two types of queries are shown.}
    \label{fig:llama2_comparison}
\end{figure*}

We observe that the LLM's representation space is highly sensitive to harmful, benign, and jailbreak attacks. First, we analyze the performance of four open-source alignment LLMs within their representation space and compared the hidden state representations of harmful and benign queries through PCA visualization, as shown in Figure~\ref{fig2}. We find that alignment LLMs can naturally distinguish between harmful and benign queries within their representation space, demonstrating that the alignment LLM's hidden states indeed have the capability to indicate the LLM's next behavior. Then, we evaluate six jailbreak attacks alongside harmful and benign queries within these alignment LLM's representation spaces, as shown in Figure \ref{fig:llama2_comparison}. The results indicate that the distribution of jailbreak attacks is closer to harmful queries than benign ones. Therefore, we propose using the LLM's hidden states to identify the LLM's harmful behavior, leveraging the LLM's inherent complex capabilities to enhance its security.

In this paper, we introduce the Hidden State Filter (HSF), a simple yet effective classification-based defense against jailbreak attacks. HSF leverages the classification abilities developed during the model's alignment phase, turning the defense against complex jailbreak attacks into a straightforward classification task without needing costly retraining or fine-tuning of the LLM. Specifically, we firstly samples features from the last k tokens in the final Decoder Layer during inference on datasets containing harmful, benign, and adversarial queries. Then, we use these features to train a lightweight classification model, integrated as a plug-in module after the final Decoder Layer. This allows for early detection of jailbreak attacks before inference, reducing computational overhead. Notably, HSF is efficient and modular, activating the model’s latent classification capabilities. It provides a dynamic defense mechanism applicable to existing LLMs without altering their core architecture or requiring significant additional resources. The plug-in design ensures versatility and easy integration into various LLM architectures with minimal configuration.

We evaluate the effectiveness, efficiency, and compatibility of HSF against six advanced jailbreak attacks based on two harmful content benchmarks using four open-source LLMs. The experimental results consistently show that HSF outperforms all baselines in defending against jailbreak attacks, with negligible computational overhead and a low false positive rate.

To summarize, our main contributions are as follows: 
\begin{itemize}
    \item We analyze the effectiveness of LLM hidden state representations in preventing jailbreak attacks, who indicates the LLM's upcoming behavior, thereby effectively preventing potential harmful outputs in a simple manner.
    
    \item We propose a jailbreak attack prevention method, which provides a dynamic defense mechanism applicable to existing LLMs without altering their core architecture or requiring significant additional resources. 
    
    \item Experiments on two benchmarks using four open-source LLMs show that HSF consistently outperforms all baselines, with minimal computational overhead and a low false positive rate.
\end{itemize}

\section{Related Work}
In the following sections, we provide an overview of the related work, beginning with a discussion of various approaches to jailbreak attacks, and then moving on to the defenses developed to counter these attacks.
\subsection{Jailbreak Attacks}
Current jailbreak attacks can be broadly categorized into two major types: template completion attacks and prompt rewriting attacks\cite{yi2024jailbreak}. 

In template completion attacks, the attacker embeds harmful queries within a contextual template to generate jailbreak prompts. \citet{liu2023jailbreaking} investigates adversarial examples against the GPT-3 model, where carefully crafted inputs induce the model to generate inappropriate content. \citet{wei2024jailbroken} identifies the root cause of LLMs' vulnerability to jailbreak attacks as conflicting objectives and generalization mismatches, noting that the model's safety capabilities do not match its complexity. \citet{li2023deepinception} constructs a virtual, nested scenario to hypnotize large models into performing jailbreak attacks. \citet{ding2024wolf} utilizes prompt rewriting and scenario nesting within LLMs themselves to generate effective jailbreak prompts. \citet{wei2023jailbreak} leverages the model’s contextual learning abilities to guide the LLM in generating unsafe outputs.

Prompt rewriting attacks involve constructing jailbreak prompts through optimization techniques, with two main approaches: (1) Gradient-based methods, as \citet{zou2023universal} uses gradient optimization to generate adversarial inputs; (2) Genetic algorithm-based methods, as \citet{liu2024autodan} and \citet{yu2023gptfuzzer} collect seeds from a curated library of handcrafted templates and continuously generate optimized adversarial prompts through mutation and crossover.

Moreover, recent advancements in red-teaming approaches further extend the capabilities of jailbreak attacks. \citet{hong2024curiosity} introduces a curiosity-driven red-teaming approach, which leverages curiosity-based reinforcement learning to explore a wide range of potential adversarial prompts. \citet{samvelyan2024rainbow} cast adversarial prompt generation as a quality-diversity problem and uses open-ended search to generate prompts that are both effective and diverse. \citet{lee2024learning} uses GFlowNet fine-tuning, followed by a secondary smoothing phase, to train the attacker model to generate diverse and effective attack prompts.

\subsection{Existing Defenses}
We categorize existing defense measures into two main types: model-level defenses and prompt-level defenses.

Model-Level Defenses: \citet{bianchi2024safety} demonstrates that incorporating appropriate safety data during the fine-tuning process of Llama can significantly enhance the model's safety. \citet{bai2022training} fine-tunes language models using preference modeling and RLHF to make them useful and harmless assistants. \citet{ouyang2022training} fine-tune GPT-3 using reinforcement learning from human feedback (RLHF), resulting in the creation of InstructGPT. \citet{ji2024aligner} designs a novel and simple alignment paradigm that learns the correctional residuals between preferred and dispreferred answers using a small model. \citet{wang2024mitigating} proposes the Backdoor Enhanced Safety Alignment method inspired by an analogy with the concept of backdoor attacks.

Prompt-Level Defenses: \citet{jain2023baseline} and \citet{alon2023detecting} employ input perplexity as a detection mechanism to defend against optimization-based attacks. \citet{phute2023llm} leverages the LLM itself to detect whether harmful content is being generated. \citet{jain2023baseline} proposes using paraphrasing and re-tokenization as defenses against optimization-based attacks, both of which involve modifying the input. \citet{xie2023defending} utilizes self-reminders in system prompts to encourage LLMs to respond responsibly, thereby reducing the success rate of jailbreak attacks. \citet{zheng2024prompt} introduces a secure prompt optimization method that shifts the representation of a query either toward the rejection direction or away from it, depending on the harmfulness of the query. Our HSF falls into this category. Compared to existing methods, HSF mitigates jailbreak attacks without compromising the quality of the model’s output or incurring additional inference costs.

\section{Motivation}
In this section, we explore the representation capabilities of LLMs in distinguishing harmful queries, benign queries, and harmful queries with adversarial prefixes.

\subsection{Data Synthesis and Basement Model}
If harmful and benign queries can be distinguished, we want this distinction to arise from their differences in harmfulness rather than other unrelated factors such as format or length. To address the impact of unrelated features, we used the commercial API of ChatGPT, gpt-3.5-turbo, to synthesize harmful and benign queries with fine control.

To ensure the diversity of the synthesized dataset, we sampled the top 100 data points from Advbench \cite{zou2023universal} as harmful queries and instructed gpt-3.5-turbo to generate benign versions of these queries while maintaining consistency in length as much as possible. Please refer to the appendix for the prompts we used to guide the data synthesis. We then applied additional manual checks to ensure effectiveness and quality. As a result, we collected 100 harmful and 100 benign queries, with an average length of 15.05 and 17.14 tokens, respectively (measured using the Llama2-7b-chat tokenizer).

We conduct experiments with three popular 7B chat LLMs available on HuggingFace, as well as an uncensored 30B LLM: Llama-2-chat \cite{touvron2023llama}, Vicuna-v1.5 \cite{chiang2023vicuna}, Mistral-instruct-v0.2 \cite{jiang2023mistral}, and WizardLM-30B-Uncensored~\cite{wizardlm30b_uncensored}. Some of these models explicitly underwent extensive safety training (Llama2-7b-chat and Mistral-instruct-v0.2).

\subsection{Visualization Analysis}
In this section, we first compare the performance of aligned and unaligned models in LLM's hidden state representations for harmful and benign queries. Then, we examine whether LLM's hidden state have the ability to distinguish between harmful, harmless, and jailbreaking.

\subsubsection{Aligned Model vs. Unaligned Model} \hfill \\
Following~\cite{zheng2024prompt}, we used Principal Component Analysis (PCA) to visualize the hidden states of the models. We selected the hidden states of the last input token from the model's last Decoder Layer output, as these hidden states intuitively gather all the information about the model’s understanding of the query and its response strategy. It should be noted that these hidden states are also projected through the language modeling head (a linear mapping) for the prediction of the next token, which means they present a linear structure in the corresponding representation space (as assumed by PCA). We use two sets of hidden states to compute the first two principal components for each model, which include both harmful and benign queries. By selecting these data points, we are able to extract the most significant features related to the harmfulness of the queries.

We find that the alignment models have a significantly better ability to distinguish between harmful and benign queries compared to unaligned models. From the top half of Figure~\ref{fig2}, it can be seen that models trained with safety measures, such as  Llama2-7B-chat, Vicuna-v1.5 and Mistral-instruct-v0.2, can naturally distinguish harmful queries from benign ones, with their boundaries (black dashed lines) easily fitted by logistic regression using the harmfulness of the queries as labels. However, the unaligned model WizardLM-30B-Uncensored lack this capability.

\subsubsection{Distinguishing Ability for Jailbreak Attack} \hfill \\
Inspired by this finding, we construct six different jailbreak attacks based on the easyjailbreak dataset and the 50 harmful queries from the aforementioned synthesized dataset. The attack methods include AutoDAN \cite{liu2024autodan}, DeepInception \cite{li2023deepinception}, GPTFuzz\cite{yu2023gptfuzzer}, ICA\cite{wei2023jailbreak}, ReNellm\cite{ding2024wolf}, and gcg\cite{zou2023universal}, each aiming to bypass the model's safety mechanisms through different approaches. For  Llama2-7B-chat, we use three sets of hidden states to compute the first two principal components, which included benign queries, harmful queries, and jailbreak attacks.

We observe that models trained with safety protocols can not only distinguish between harmful and benign queries but also differentiate jailbreak attacks. The boundaries between these categories (indicated by black dashed lines) can be easily fitted using an SVM, with the harmfulness of the queries still serving as the labels. More reuslts can be found in Appendix\ref{appendix:more results}. Based on this observation, our insights for developing a lightweight classification model include: (i) employing more robust methods for sampling the classification criteria, and (ii) more accurately fitting the boundaries between benign queries, harmful queries, and jailbreak attacks within the model's representation space.
\section{Hidden State Filter}
In this section, we present an overview of the HSF, followed by a detailed description of its design.

\subsection{Overview of Hidden State Filter}
\begin{figure*}[!t] 
    \centering
    \includegraphics[width=13cm]{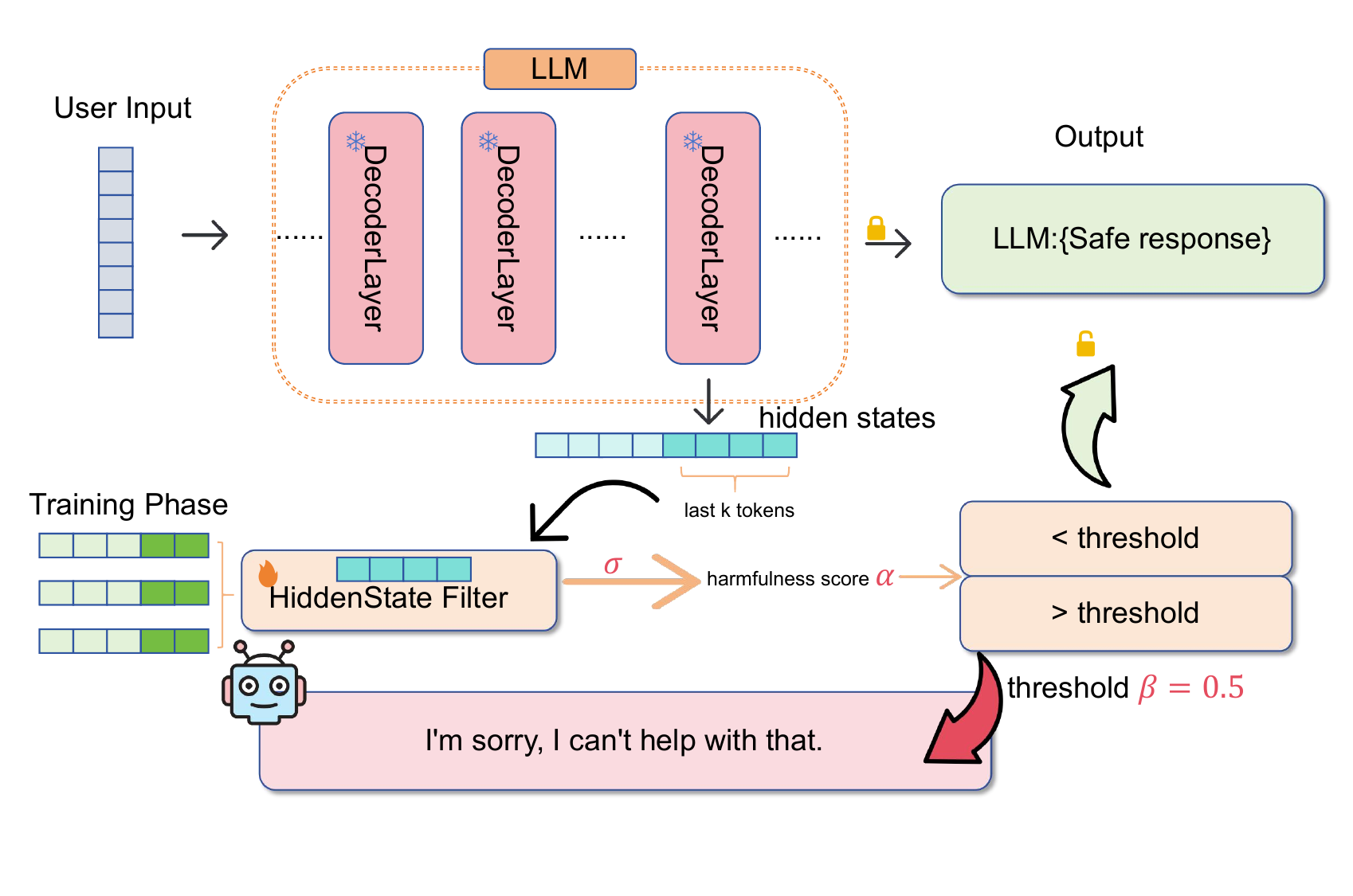} 
    \caption{Schematic representation of the HSF mechanism. In this process, the Large Language Model (LLM) forwards the hidden states from the final Decoder Layer to the trained HSF. The filter extracts the last \(k\) tokens, concatenates them, and computes a harmfulness score \(\alpha\) using a sigmoid function. This score is then compared to a threshold to determine the model's response, ensuring safe and reliable output.}
    \label{fig:outline}
\end{figure*}

Our HSF consists of two stages, as illustrated in Figure\ref{fig:outline} . The first stage is the training phase, where a weak classifier is constructed to classify user queries. This weak classifier is trained using hidden vectors extracted from the last Decoder Layer during the forward pass of a specified LLM. In the second inference stage, the user's query undergoes a forward pass through the LLM to obtain the hidden vectors from the last Decoder Layer. HSF then classifies the last \( k \) tokens of these hidden vectors to determine whether to block the current inference. The remainder of this section details each step. It is important to note that the HSF does not generalize across different models; therefore, for clarity, we will illustrate the process using Llama2-7B-chat  and Mistral-instruct-v0.2 as the example model.

\subsection{Training Phase: A Weak Classifier}
To construct the expert model, we first collect 3,000 samples each from the UltraSafety \cite{guo2024controllable} and PKU-SafeRLHF-prompt \cite{ji2024beavertails} datasets as the harmful query dataset. These queries are expected to be rejected by any LLM aligned with human values. We also collect 6,000 samples from the databricks-dolly-15k \cite{DatabricksBlog2023DollyV2} dataset as the benign query dataset. For the poorly aligned mistral-instruct-v0.2 model, we additionally sample 750 queries from both the harmful and benign datasets we construct. These samples are used to create a positive and negative dataset for training by adding template completion jailbreak attack templates.

Step 1: To construct the training samples, we input these queries into the model and perform a forward pass, extracting the vectors corresponding to the last \( k \) tokens from the hidden state of the last Decoder Layer, denoted as \( t_{k} \in \mathbb{R}^{n\times k} \). Since the default chat template is used when inputting queries into the LLM, the token length is at least 8 for the Llama2-7B-chat tokenizer. Therefore, we set the search space for the last \( k \) tokens \( t_{k} \) as \( k \in \{1, 2, \ldots, 8\} \).

Step 2: Concatenation of \( k \) vectors. The vectors corresponding to the last \( k \) tokens are concatenated in their original order, with zeros padded between every two tokens to soft-distinguish different tokens and optimize the classifier’s recognition capability. The specific formula is as follows:

Let \( t_1, t_2, \ldots, t_k \) be the vectors corresponding to the last \( k \) tokens, where \( t_i \in \mathbb{R}^n \). We define the vector \( T_k \) as:

\begin{equation}
T_k = [t_k, \mathbf{0}, t_2, \mathbf{0}, \ldots, \mathbf{0}, t_1],
\end{equation}

\noindent where \( \mathbf{0} \in \mathbb{R}^n \) denotes a zero vector of length \( n \).

Step 3: Constructing the weak classifier. Considering the model alignment capability extension and inference time cost control, we use a simple Multilayer Perceptron (MLP) as the classifier. We fit a logistic regression model using the empirical rejection probability of \( T_k \), applying dropout regularization to prevent overfitting:

\begin{equation}
f_k : \mathbb{R}^{n\times k} \to \mathbb{R}, \quad f_k(x) = w_k^T T_k + b_k,
\end{equation}

\noindent where \( w_k \in \mathbb{R}^{n \times k} \) and \( b_k \in \mathbb{R} \) are the parameters to be fitted by the logistic regression model.

Step 4: Model training. We train the model by minimizing the binary cross-entropy loss function. The specific training steps include forward propagation to compute the loss and backpropagation to update the parameters.

For a given query sample \( T_k \), we define the following binary cross-entropy loss function:

\begin{equation}
\mathcal{L}_k(\theta) = -l \log(\sigma(f_k(T_k))) - (1 - l) \log(1 - \sigma(f_k(T_k))),
\end{equation}

\noindent where:
 \( l \in \{0, 1\} \) is the binary label indicating whether the query is safe,
 \( \sigma \) is the sigmoid function.

To optimize the model parameters \( \theta \), we minimize the above loss function \( \mathcal{L}_r(\theta) \), ensuring that harmful queries \( (l=1) \) receive a higher harmfulness score and benign queries \( (l=0) \) receive a lower harmfulness score.

\begin{table*}[!t]
    \centering
    \small
    \caption{Comparison of Attack Success Rates between HSF and baseline defenses on Mistral-instruct-v0.2 and Llama2-7B-chat.}
    \begin{tabular}{llcccccccc}
        \toprule
        \multirow{2}{*}{Model} & \multirow{2}{*}{Defense} & \multicolumn{2}{c}{Harmful Benchmark $\downarrow$} & \multicolumn{6}{c}{Jailbreak Attacks $\downarrow$} \\
        \cmidrule(r){3-4} \cmidrule(r){5-10}
        & & AdvBench & MaliciousInstruct & GCG & AutoDAN & ReNellm & DeepInception & ICA & GPTFuzz \\
        \midrule
        \multirow{8}{*}{Mistral} 
        & No Defense & 77\% & 82\% & 80\% & 98\% & 84\% & 67\% & 56\% & 88\% \\
        & PPL & 77\% & 82\% & \textbf{0}\% & 98\% & 84\% & 67\% & 56\% & 88\% \\
        & Paraphrase & 34\% & 45\% & 70\% & 44\% & 47\% & 15\% & 71\% & 76\% \\
        & Retokenization & 65\% & 54\% & 70\% & 72\% & 47\% & 48\% & 82\% & 58\% \\
        & Self-Examination & 13\% & 4\% & 6\% & 19\% & 12\% & 2\% & 3\% & 28\% \\
        & Self-Reminder & 1\% & 2\% & 6\% & 92\% & 59\% & 19\% & 0\% & 80\% \\
        & DRO & 73\% & 78\% & 76\% & 100\% & 87\% & 71\% & 63\% & 93\% \\
        & SPD & 99\% & 60\% & 75\% & 4\% & 61\% & 83\% & 0\% & 77\% \\
        & HSF & \textbf{0\%} & \textbf{1\%} & 4\% & \textbf{0\%} & \textbf{8\%} & \textbf{0\%} & \textbf{0\%} & \textbf{0\%} \\
        \midrule
        \multirow{8}{*}{Llama2} & No Defense & 0\% & 0\% & 38\% & 10\% & 20\% & 44\% & 0\% & 30\% \\
        & PPL & 0\% & 0\% & \textbf{0\%} & 10\% & 20\% & 44\% & 0\% & 30\% \\
        & Paraphrase  & 0\% & 1\% & 4\% & 9\% & 11\% & 29\% & 0\% & 29\% \\
        & Retokenization & 6\% & 5\% & 9\% & 19\% & 49\% & 53\% & 0\% & 31\% \\
        & Self-Examination & 0\% & 0\% & 6\% & 0\% & 0\% & 0\% & 0\% & 2\% \\
        & Self-Reminder & 0\% & 0\% & 0\% & 4\% & 0\% & 0\% & 0\% & 2\% \\
        & DRO & 0\% & 0\% & 2\% & 9\% & 22\% & 32\% & 0\% & 30\% \\
        & SPD & 0\% & 0\% & 100\% & 12\% & 61\% & 100\% & 0\% & 2\% \\
        & HSF & \textbf{0\%} & \textbf{0\%} & 1\% & \textbf{0\%} & \textbf{0\%} & \textbf{0\%} & \textbf{0\%} & \textbf{1\%} \\
        \bottomrule
    \end{tabular}

    \label{ASR}
\end{table*}

\subsection{Defense Phase: Implementing Jailbreak Attack Defense by Classifying User Queries}
In the defense phase, the trained weak classifier is inserted into LLM as a plug-in module. The process is as follows:

1. Receive the concatenated vector \( T_k \) of the selected last \( k \) tokens as input.

2. Input the concatenated vector \( T_k \) into the classifier to calculate a harmfulness score \( \alpha \) (ranging from 0 to 1).

3. Specify a threshold \( \beta \); if the harmfulness score exceeds this threshold, the inference process is interrupted, and the model refuses to generate a response.

\section{Experiments}
In this section, we evaluate the defense performance of the HSF, focusing on two key metrics: Attack Success Rate (ASR) and Area Under the Curve (AUC).

\subsection{Experimental Setup}

\textbf{Basement Models.} We deploy HSF on three open-source LLMs: Vicuna-v1.5 \cite{chiang2023vicuna}, Llama2-7b-chat \cite{touvron2023llama}, Mistral-instruct-v0.2 \cite{jiang2023mistral}to evaluate its effectiveness.

\noindent\textbf{Attack Methods.} We consider six state-of-the-art jailbreak attacks, each covering different categories, implemented through easyjailbreak \cite{zhou2024easyjailbreak}. Specifically, GCG \cite{zou2023universal} represents gradient-based attacks, AutoDAN \cite{liu2024autodan} is based on genetic algorithms, and ICA \cite{wei2023jailbreak}represents context-learning-based attacks. We also included ReNellm \cite{ding2024wolf}, DeepInception \cite{li2023deepinception}, and GPTFuzz \cite{yu2023gptfuzzer} as representative empirical jailbreak attacks. To evaluate the defense performance against naive attackers directly inputting harmful queries, we used two harmful query benchmark datasets: Advbench \cite{zou2023universal} and MaliciousInstruct \cite{huang2024catastrophic}. Detailed settings for these attack methods and harmful query datasets can be found in the Appendix\ref{appendix:attack setup}.

\noindent\textbf{Baselines.} We consider six state-of-the-art defense mechanisms as baselines. PPL\cite{alon2023detecting} and Self-Examination \cite{phute2023llm}) are input-output detection-based methods, while Paraphrase \cite{jain2023baseline}, Retokenization \cite{jain2023baseline}, and Self-Reminder \cite{xie2023defending} are mitigation-based methods. DRO\cite{zheng2024prompt} is a safety-prompt-based method. SPD\cite{candogan2024single} is a method for detecting jailbreaking inputs via the logit values in a single forward pass.Note that the original classifier used in SPD is SVM. However, due to issues encountered during reproduction, it has been replaced with an equivalent MLP structure in this study.Detailed descriptions and hyperparameter settings for each method can be found in the Appendix\ref{appendix:attack setup}.

\noindent\textbf{Evaluation Metrics.} We utilize Llama-Guard-3-8B \cite{dubey2024llama3herdmodels} to evaluate the safety of the model's responses. Although keyword-based detection methods \cite{zou2023universal} are commonly employed, our experiments revealed that these methods had a low AUC, indicating a high false positive rate. This necessitated additional manual verification, leading to increased costs. Consequently, we employ the state-of-the-art Llama-Guard-3-8B to assess whether the model generated unsafe content and to calculate the ASR.

\noindent\textbf{HSF Settings.} We set the hyperparameter \(k=7\) for Llama2-7B-chat, meaning that we used the last 7 tokens of the last DecoderLayer as the basis for the HSF input. We also set the harmfulness detection threshold to 0.5. In section \ref{ablation},We present an ablation analysis of different values of the last \(k\) tokens.

\subsection{Experimental Results}

\textbf{HSF Enhances LLM Safety.} Table \ref{ASR} presents a comparison of the ASR for the Mistral-instruct-v0.2 and Llama2-7B-chat models when employing HSF and various baseline defenses against six different jailbreak attacks. The results indicate that for models with weaker safety alignment, such as Mistral-instruct-v0.2, HSF markedly decreases the ASR, outperforming almost all baseline defenses. Notably, while most other defenses were ineffective against AutoDAN, HSF successfully defends against this attack, achieving an ASR of 0. For models with strong alignment, such as Llama2-7B-chat, HSF consistently reduces the ASR across all attacks to nearly 0. Additional results for HSF applied to Vicuna-v1.5 can be found in Appendix\ref{appendix:more results}. 

Moreover, HSF has a high AUC score. Table \ref{AUC} shows the AUC score of HSF on safe datasets.

\begin{table*}[!ht]
    \centering
    \renewcommand{\arraystretch}{1.2}  
    \caption{Comparison of AUC for Mistral-instruct-v0.2 and Llama2-7B-chat across different K values (1-8).}
    \begin{tabular}{lccccccccc}
        \toprule
        \multicolumn{1}{c}{} & \multicolumn{8}{c}{AUC $\uparrow$} \\
        \cmidrule(lr){2-9}
        Model & K=1 & K=2 & K=3 & K=4 & K=5 & K=6 & K=7 & K=8 \\
        \midrule
        Mistral & 0.9990 & 0.9986 & \textbf{0.9998} & 0.9991 & 0.9990 & 0.9992 & 0.9997 & 0.9988\\
        \midrule
        Llama2 & \textbf{0.9999} & 0.9994 & 0.9994 & 0.9998 & \textbf{0.9999} & 0.9997 & \textbf{0.9999} & 0.9997 \\
        \bottomrule
    \end{tabular}

    \label{AUC}
\end{table*}

\subsection{Ablation Analysis}
\label{ablation}

\begin{table*}[!ht]
    \centering
    \caption{Ablation analysis on the hyperparameter \(k\) for HSF.}
    \begin{tabular}{llcccccccc}
        \toprule
        \multirow{2}{*}{Model} & \multirow{2}{*}{Last K} & \multicolumn{2}{c}{Harmful Benchmark $\downarrow$} & \multicolumn{6}{c}{Jailbreak Attacks $\downarrow$} \\
        \cmidrule(r){3-4} \cmidrule(r){5-10}
        & & AdvBench & MaliciousInstruct & GCG & AutoDAN & ReNellm & DeepInception & ICA & GPTFuzz \\
        \midrule
        \multirow{8}{*}{Mistral} & k=1 & 1\% & 1\% & 16\% & 0\% & 3\% & 6\% & 0\% & 0\% \\
        & k=2 & 1\% & 0\% & 4\% & 0\% & 12\% & 0\% & 0\% & 0\% \\
        & \textbf{k=3}  & 0\% & 1\% & 4\% & 0\% & 8\% & 0\% & 0\% & 0\% \\
        & k=4 & 1\% & 0\% & 4\% & 0\% & 17\% & 0\% & 0\% & 0\% \\
        & k=5 & 0\% & 1\% & 4\% & 0\% & 66\% & 0\% & 0\% & 0\% \\
        & k=6 & 0\% & 2\% & 6\% & 0\% & 64\% & 0\% & 0\% & 0\% \\
        & k=7 & 0\% & 0\% & 6\% & 0\% & 62\% & 0\% & 0\% & 0\% \\
        & k=8 & 0\% & 0\% & 2\% & 0\% & 66\% & 0\% & 0\% & 0\% \\
        \midrule
        \multirow{8}{*}{Llama2} & k=1 & 0\% & 1\% & 21\% & 0\% & 23\% & 48\% & 0\% & 1\% \\
        & k=2 & 0\% & 0\% & 8\% & 0\% & 21\% & 50\% & 0\% & 1\% \\
        & k=3  & 1\% & 0\% & 1\% & 0\% & 23\% & 12\% & 0\% & 1\% \\
        & k=4 & 1\% & 0\% & 3\% & 0\% & 21\% & 3\% & 0\% & 1\% \\
        & k=5 & 0\% & 0\% & 2\% & 0\% & 21\% & 58\% & 0\% & 0\% \\
        & k=6 & 0\% & 0\% & 1\% & 0\% & 29\% & 60\% & 0\% & 1\% \\
        & \textbf{k=7} & 0\% & 0\% & 1\% & 0\% & 0\% & 0\% & 0\% & 1\% \\
        & k=8 & 0\% & 0\% & 1\% & 0\% & 31\% & 66\% & 0\% & 1\% \\
        \bottomrule
    \end{tabular}

    \label{Ablation}
\end{table*}

In this section, we conduct an ablation analysis on the hyperparameter \(k\) in HSF, tested on Mistral-instruct-v0.2 and Llama2-7B-chat models. Table \ref{Ablation} show that HSFr exhibits significant robustness against prompt rewriting attacks, maintaining an extremely low ASR regardless of the \(k\) value. However, for more complex template completion attacks, such as ReNellm and DeepInception, the defense performance of the model shows a clear dependency on the hyperparameter settings. This suggests that to maintain high classification performance in advanced attack scenarios, HSF requires appropriate hyperparameter tuning.

Moreover, we find that due to the identical chat template used, the last 4 tokens of the model input remained consistent. As a result, HSF exhibites a lower ASR at \(k=4\), demonstrating better defense effectiveness. Additionally, for GCG jailbreak attacks, when \(k > 1\), HSF effectively disrupts its robustness, leading to a significant drop in ASR.

\section{Limitations}
\textbf{Limited Generalizability}: The HSF is specifically tailored to work with certain LLMs, necessitating retraining when applied to other models. This dependency on the alignment capabilities of the underlying model restricts its generalizability to models that have undergone analogous alignment training. However, while retraining is required for each new model, the process incurs relatively low computational cost and does not alter the model’s architecture.
\vfill\eject

\section{Conclusion}
In this paper, we introduce a novel model architecture module, HSF, designed to defend against jailbreak attacks on LLMs by leveraging the model's alignment capabilities reflected in its hidden states. We observe that aligned models can internally distinguish between jailbreak attacks, harmful queries, and benign queries. Inspired by this observation, we develope a plug-in weak classifier module that uses the last \(k\) tokens from the LLM's DecoderLayer to maximize the utilization of the model's complex alignment capabilities as a safety measure. This ability allows HSF to identify the harmfulness of queries before inference begins. Our results demonstrate that HSF can effectively defend against state-of-the-art jailbreak attacks while being both efficient and beneficial. In future, we will explore how to leverage LLM hidden states to build more flexible harmful-behavior detectors and enhance jailbreak attack prevention through Retrieval-Augmented Generation methods.

\section*{Acknowledgments}
This work was funded by the National Natural Science Foundation of China (NSFC) under Grants No. 62406013, the  Beijing Advanced Innovation Center Funds for Future Blockchain and Privacy Computing and the Fundamental Research Funds for the Central Universities. 
\clearpage
\bibliographystyle{ACM-Reference-Format}
\balance
\bibliography{wsai25}

\appendix
\section{Appendix A: Detailed Experimental Setups}

\subsection{Attack Setup}
\label{appendix:attack setup}
For GCG \cite{zou2023universal}, AutoDAN \cite{liu2024autodan}, GPTFuzz \cite{yu2023gptfuzzer}, DeepInception \cite{li2023deepinception}, and ICA \cite{wei2023jailbreak}, we use the easyjailbreak \cite{zhou2024easyjailbreak} framework to construct attacks. We utilized 100 different representative harmful queries from Advbench \cite{zou2023universal}. For ReNellm \cite{ding2024wolf}, we rewrite these 100 representative harmful queries collected from Advbench using easyjailbreak and filled them into templates to create 100 jailbreak prompts.

\subsection{Baseline Setup}

\textbf{1. PPL \cite{alon2023detecting}}: PPL is an input detection mechanism that calculates the perplexity of a given input to determine whether the user's request should be accepted or rejected.

The definition of perplexity (P) is as follows:
\begin{equation}
\text{Perplexity}(P) = \exp\left(\frac{-1}{N} \sum_{i=1}^{N} \log P(w_i \mid w_{1:i-1})\right).
\end{equation}
Following \cite{alon2023detecting}, we use Llama2-7b-chat to calculate perplexity. As per \cite{xu2024safedecoding}, we set the PPL threshold to the highest perplexity value of harmful queries in Advbench. This ensures that queries from Advbench do not trigger the detector.

\noindent\textbf{2. Paraphrase \cite{jain2023baseline}}: We follow \cite{jain2023baseline} and used GPT-4o by default to paraphrase user queries. The prompt is as follows:

The paraphrased output is then used as the input to the target language model.

\begin{figure}[!ht]
    \centering
    \includegraphics[width=\columnwidth]{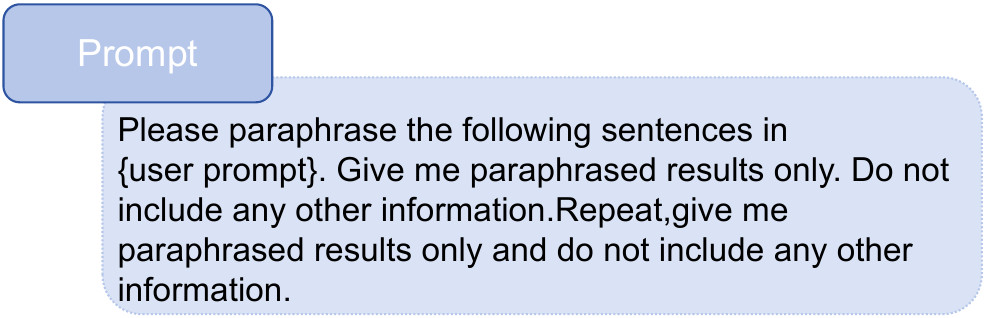}
   
\end{figure}

\noindent\textbf{3. Retokenization \cite{jain2023baseline}}: This method splits tokens and represents them using multiple smaller tokens. For example, the token representing the word “breaking” is split into two tokens representing “break” and “ing.” We use BPE-dropout \cite{provilkov2020bpe} , which randomly drops p\% of BPE merges during the tokenization process. We set p = 0.2 following \cite{jain2023baseline}.

\noindent\textbf{4. Self-Examination \cite{phute2023llm}}: This is an output detection method that uses the language model itself to distinguish whether harmful content is being generated. We use the prompt suggested by \cite{phute2023llm}:
\begin{figure}[!ht]
    \centering
    \includegraphics[width=\columnwidth]{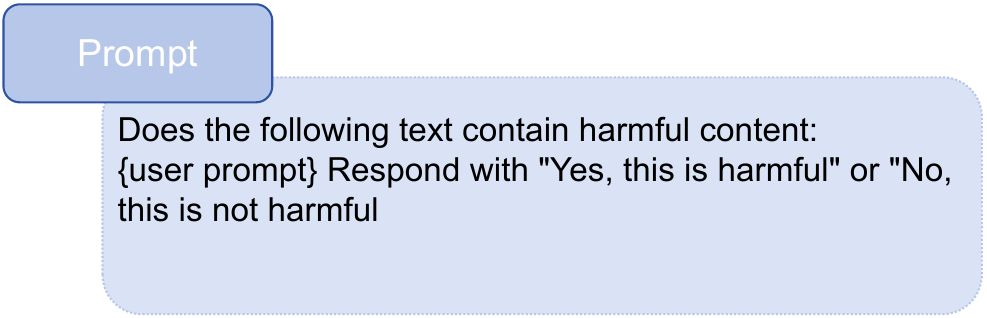}
   
\end{figure}

\noindent\textbf{5. Self-Reminder\cite{xie2023defending}}: Self-Reminder appends prompts to input prompts to remind the language model to respond responsibly.
\begin{figure}[!ht]
    \centering
    \includegraphics[width=\columnwidth]{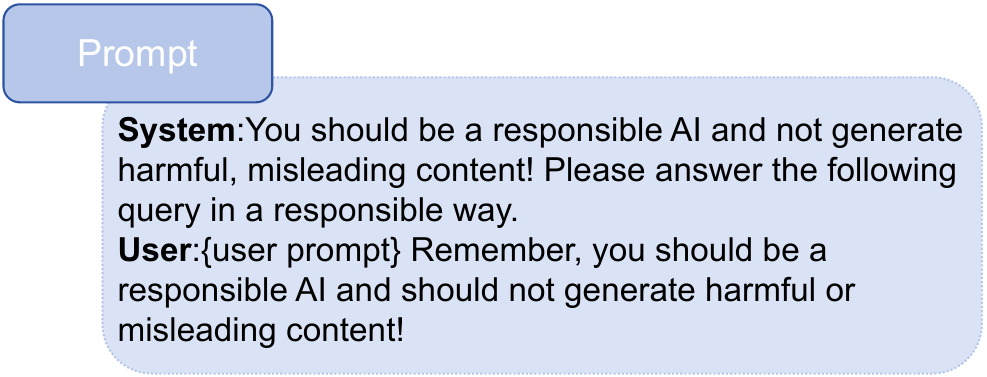}
   
\end{figure}

\noindent\textbf{6. DRO \cite{zheng2024prompt}}: This is an automated optimization method for generating safe prompts. Following previous work, we use a default type of system safety prompt and directly embedded the generated safety prompt as a vector in the embedding layer before the user's input.

\noindent\textbf{7. SPD \cite{candogan2024single}}: This is a method to detect adversarial samples by extracting a feature matrix \( H \) from the logit outputs of a LLM. \( H \) captures unnatural token distribution changes caused by attacks, using the top-\( k \) logits for \( r \) token positions.The authors originally used a Support Vector Machine (SVM) with an RBF kernel trained on \( H \) for classification. However, during our reproduction, we encountered issues with the SVM and instead replaced it with a MLP. The hyperparameters include \( r \) (number of token positions, set to 5), \( k \) (number of top logits, set to 50).

\end{document}